# A light-stimulated neuromorphic device based on graphene hybrid phototransistor

Shuchao Qin[#], Fengqiu Wang[#]*, Yujie Liu, Qing Wan, Xinran Wang, Yongbing Xu, Yi Shi, Xiaomu Wang*, Rong Zhang*

School of Electronic Science and Engineering and Collaborative Innovation Center of Advanced Microstructures, Nanjing University, Nanjing 210093, China

**Neuromorphic chip refers to an unconventional computing architecture that is modelled on biological brains[1-3]. It is ideally suited for processing sensory data for intelligence computing, decision-making or context cognition. Despite rapid development, conventional artificial synapses[4-12] exhibit poor connection flexibility and require separate data acquisition circuitry, resulting in limited functionalities and significant hardware redundancy. Here we report a novel light-stimulated artificial synapse based on a graphene-nanotube hybrid phototransistor that can directly convert optical stimuli into a "neural image" for further neuronal analysis. Our optically-driven synapses involve multiple steps of plasticity mechanisms and importantly exhibit flexible tuning of both short- and long-term plasticity. Furthermore, our neuromorphic phototransistor can take multiple pre-synaptic light stimuli via wavelength-division multiplexing and allows advanced optical processing through charge-trap-mediated optical coupling. The capability of complex neuromorphic functionalities in a simple silicon-compatible device paves the way for novel neuromorphic computing architectures involving photonics[13].**

Inspired by biological neural systems, neuromorphic chips are rapidly developed as a viable technological avenue in artificial intelligence. In stark contrast to traditional von Neumann computers, neuromorphic devices are dedicated to processing data and interacting with the world in humanlike ways[1, 11]. This manner renders neuromorphic chips extremely effective for solving complex tasks such as image recognition, multi-object detection and visual signal classification, which are beyond the capabilities of conventional semiconductor devices. In biological neural systems, synapses whose connectivity response depends on the history of stimuli previously experienced[14], act as the most fundamental computing element. The changing of connectivity, also known as synaptic plasticity, is responsible for both short- and long-term memory behaviors, and the assembly of synapses produces functionally significant operations[15]. Stimulated by such biological systems, several artificial synaptic devices that may potentially meet the scalability requirements have been developed based on either transistors[5-11] or memorisistors[16-19].

Despite dramatic advancement, state-of-the-art synaptic devices with pure electronic components present two major limitations. First, in most conventional artificial synapses, the neuromorphic computing is isolated from the data acquisition sensors (ocular, olfactory or auditory stimuli)[20, 21]. The lack of neuromorphic sensing results in huge hardware redundancy and system latency. Furthermore, real neuronal system always involves multiple steps of plasticity mechanism that enable considerable flexibility in the modulation of the connectivity strength[14, 22, 23]. For a given artificial synaptic pair, the coupling coefficient of these devices is always fixed, which is not adequate to emulate the complex activities of an organism. In short, the *in-situ* adjustment of synaptic weight is fundamental to highly complex behaviors of neural activities yet has remained a significant challenge for current synaptic technologies[24]. These limitations are calling for new building blocks and architectures for the next-generation neuromorphic systems.

Here, we demonstrate a proof-of-concept light-stimulated neuromorphic device based on a phototransistor combining atomically thin graphene with single-walled carbon nanotubes (SWNTs). In our device, short-term plasticity (STP) behaviors, including paired-pulse facilitation and spatiotemporally correlated elementary dynamic logic, are implemented through charge transfer between graphene and SWNTs[25, 26]. The plasticity can be flexibly modulated by the gate voltage, which results in a dynamic synapse with adjustable weight. In addition, due to the charge-trap rich interface between the channel and the substrate, the synapse can also act as an optoelectronic non-volatile memory that allows gate-controlled rewriting. This light-enhanced memory behavior well emulates long-term plasticity (LTP) in synapses. Combining these features, our device naturally works as an artificial retina that is fundamentally different from existing architectures based on non-neuromorphic sensors. Furthermore, by exciting the device with multiple optical stimuli, we demonstrate the capability of parallel and complex optical neuromorphic processing, a feature highly desirable for future brain-inspired computation.

The schematic of the proof-of-concept light-stimulated synapse is illustrated in Fig. 1a. Details of device fabrication are reported in Ref. [25]. Compared with conventional electronic devices, the utilization of light as stimuli provides a number of technological advantages. For example, lightwaves generally offer much broader bandwidth and much faster processing speeds than electronics[13]. Intrinsic light degree-of-freedom such as state of polarization may be exploited for emulating complex neural activities. In our synaptic transistor, the light pulses irradiated onto the device generate photocarriers that alter the channel conductance. Therefore, the light pulse can be regarded as the pre-synaptic spikes or external stimuli, and the channel conductance is treated as the synaptic weight. The tunable transport properties of channel material are the foundation of the dynamically tunable synaptic plasticity[24]. Here, the gate bias modulates the carries transportation, thereby achieving an *in-situ* adjustment of synaptic strength.

We first characterize the light-activated short-term plasticity in our device. When a pre-synaptic light spike (405 nm, 50 μW, 5 ms) is applied, a typical inhibitory post-synaptic current (IPSC) is directly triggered at $V_G$=0 V. It reaches a peak value immediately after the excitation spike, and gradually recovers to its initial value (top inset of Fig. 1b). This dynamical behavior well reproduces an IPSC process observed in biological inhibitory synapse[27] and is related to the STP mechanism. Changing the gate bias to $V_G$=20 V, the same input light spike triggers an excitatory post-synaptic current (EPSC), which matches the behavior of an excitatory synapse[21] in organisms (the bottom inset of Fig. 1b). The recovering channel conductance can be fitted with a double exponential function. The two decay times may be attributed to the spatial separation of photoexcited charge carriers (Supplementary Section S1). These synaptic responses are highly stable and can always be observed after the applications of input light pulses (Supplementary Fig. S1). As the ΔPSC (difference between initial and peak value of post-synaptic current) is an indication of the synaptic plasticity, these results prove that the plasticity of the artificial synapse can be modulated by the gate electric field. To further demonstrate *in-situ* adjustment of the synaptic weight, the ΔPSCs triggered by the spike under different gate biases from -50 V to +50 V are studied, as plotted in Fig. 1b. The saturation of ΔPSCs is attributed to the high density of states of graphene, *i.e.* the Fermi level of graphene is not as amendable at high gate voltage. Obviously, the gate-controllable ΔPSC represents a highly flexible and convenient plasticity modulation of the synapses.

In neuroscience, spike-timing-dependent plasticity (STDP) is widely accepted as a synaptic learning rule, where the synaptic modification relies on relative timing of neuronal activity[28]. The spike duration-dependent inhibitory or excitatory post-synaptic current variations of our device are shown in Fig. 1c. The amplitude of ΔIPSC or ΔEPSC peak values increase with the spike duration from 5 to 100 ms. The saturated increase of the ΔPSC is attributable to the photon absorption saturation[29]. It should be mentioned that the effective spike duration is very similar to those find in biological neurons [12, 30].

The inset of Fig. 1c shows a real optical microscope image of the synaptic transistor. The working principle of the synapse with adjustable plasticity can be understood through the energy band diagram in Fig. 1d. In dark condition, the work function of constituent semiconducting SWNTs is larger than that of graphene[25, 26], so the energy band tilts toward the graphene layer in order to equilibrate the Fermi level. Under light illumination, photoexcited electrons transfer from SWNTs to graphene under the built-in field, leading to a left-shift of the Dirac point (Supplementary Fig. S2). A crossover ($V_{cross}$) of the transfer curves in dark and in light illumination is clearly observed, at which point (gate voltage) the hole carrier density (more accurately, weighted by hole mobility) in dark is equivalent to the electron carrier density (weighted by electron mobility) under illumination (see Supplementary Section S2 for details). For $V_G<V_{cross}$, light illumination results in a decrease in the current, which is corresponding to an inhibitory post-synaptic behavior, and vice versa.

We now move to discuss the dynamic synaptic behaviors of our device. Paired-pulse facilitation (PPF) refers to a dynamic enhancement of transmitter release and is considered crucial for the execution of critical computation as well as for temporal information encoding in auditory or visual signals. It plays an important role in associative learning in biological systems[20]. In our synaptic transistor, this temporal learning function has also been realized. Figure 2a schematically illustrates the effect of PPF in a biological synapse. When a pair of light spikes (50 μW, 5 ms) with a pulse interval of 55 ms are applied to an inhibitory synapse ($V_G$=0 V), the peak value of the IPSC triggered by the second pre-synaptic spike is clearly larger than that by the first pre-synaptic spike, as shown in Fig. 2b. The PPF index (A2/A1×100%), defined as the ratio of the amplitude of the second IPSC (A2) divided by that of the first IPSC (A1), is plotted as a function of $\Delta t_{pre}$ (defined as the interval between the two consecutive pre-synaptic spikes) in Fig. 2c. The PPF index reaches the maximum value (155%) when $\Delta t_{pre}$ = 6 ms, and gradually decreases with increasing $\Delta t_{pre}$. These results verify that high-repetition training pulses extensively enhance learning effect in our device, which is again similar to the scenario in a real biologic system. The working mechanism relates to the recombination of photocarriers. When the first light spike ends, photo-generated electron-hole pairs begin to recombine. If the second spike is applied before the full recombination, the photo-excited electrons from the first spike still partially reside in the graphene channel. Consequently, photo-generated electrons triggered by the second light pulse are augmented, inducing the PPF. Actually, the PPF function can also be obtained in an excitatory synapse ($V_G$=20 V, see Supplementary Fig. S3). Such results demonstrate that the synaptic weight is facilitated when the interval is shorter than the recovery time of the photocarries.

Repetitive spike activity can induce a decrease or increase of synaptic efficacy in various parts of the nervous system, known as the synaptic depression or potentiation[30]. To characterize variations of the synaptic weight (defined by $\Delta$PSC/initial PSC), the trends of PSC with a series of consecutive pulses (50 μW, 5 ms pulses with 10 ms intervals) at $V_G$=0 and 20 V are measured and shown in Fig. 2d. For $V_G$=0 V, the synaptic depression characteristics (weakening of the synaptic weight) is obtained. When the light pulses are applied at $V_G$=20 V, the synaptic potentiation characteristics (strengthening of the synaptic weight) is observed. In both cases, the PSC rapidly changes in the first few light spike cycles and then saturates. These results suggest that our synaptic devices are capable of emulating biological synapses with properly designed neuron components, which provides local programming through the control of number of excitation pulses.

Multiple categories of synaptic plasticity allow synapses to perform wildly different functions in information processing[15]. We next demonstrate LTP in our device, which provides a physiological platform for learning and memory in the brain. Note that in real biological systems, distinct biochemical mechanisms from those responsible for STP should be involved in the long-term changing of connectivity. In our device, thanks to the geometry of the graphene-SWNT membrane, effects from substrate surface traps can be effectively harnessed to emulate LTP.

In Fig. 3a we show the variation of synaptic weight triggered by a long-duration pre-synaptic light spike (50 μW, 100 ms) under a negative gate bias. An accessional resistance is directly triggered but does not relax back to its initial state at $V_G$=-20 V, signifying a LTP formation. Meanwhile, LTP may be mitigated by applying a -10 V (or zero) gate voltage. Its physical mechanism again originates from photo-gating effect. Specifically, photo-generated carriers are captured by trap centers (such as dangling bonds, intrinsic defects and local structural distortions, Ref. [31]) at the substrate surface. Fig. 3b schematically illustrates this charge trapping process. Under a negative gate electric field, photo-generated holes in graphene or SWNTs are partially trapped into the trap sites, which results in a long-term stable photo-gating effect even after the incident light is switched off due to the high trapping energy barrier. The LTP memory level is determined by the trap density. Tuning the number of the traps by gate electric field effectively adjusts the LTP levels (Fig. 3a). Electrical field controlled charging/discharging of traps as well as the hysteresis loop direction of the transfer characteristics verify that the trap centers are of Coulomb type (Supplementary Fig. S4). To further examine the role of light pulse played in the LTP process, we measured the response at different illumination power (Supplementary Section S4 and Supplementary Fig. S5). With increasing incident light power, photo-induced carrier concentration rapidly increases. As a result, more carriers are trapped, leading to enhanced hysteresis window in the transfer characteristics. This phenomenon consistently supports the view that photo-excitation and gate voltage are both responsible for the memory effect observed.

Figure 3c displays the synaptic response upon consecutive spike irradiation, under a gate voltage of -20 V and -30 V, respectively. Clearly, enhancement of long-term memory, as observed in biological system[23], is emulated. In Fig. 3d, nulling of the plasticity current by a positive 'reset' gate pulse is observed. This reversed electric pulse could redistribute the trap centers and facilitate de-trapping process. These results demonstrate that our synapses can go beyond simply mimicking synaptic biological characteristics, and may also be designed as rewriteable non-volatile optical memory cells[32] or optoelectronic switches, thereby enabling more device functionalities.

We then demonstrate the possibility of emulating axon-multi-synapses networks[33] based on our light-stimulated devices. Figure 4a schematically illustrates the structure of the biological and artificial synaptic networks. In our setup, each wavelength can be coded as a pre-synaptic channel. Benefiting from the broadband light response of the graphene-SWNTs hybrid film, our device supports extremely broad bandwidth covering ultraviolet to infrared[25]. Spatiotemporally correlated photoresponse under multiple laser spikes is first revealed. Essentially, we find that our device obeys two summation rules, namely super-linear temporal summation rule and sub-linear power summation rule[34]. On one hand, we test the photo-response of the device stimulated by two laser pulses (405 nm and 532 nm light pulses, defined as pre-synapse1 and pre-synapse2) at $V_G$=0 V. The sum IPSC current, measured at the trailing edge of pre-synapse1 spike, as a function of the delay $\Delta t_{pre2-pre1}$ is plotted in Fig. 4b. When $\Delta t_{pre2-pre1}$=0, the two pre-synaptic spikes are applied simultaneously where a pronounced total $\Delta$PSC appears in the post-synaptic neuron. With increasing $|\Delta t_{pre2-pre1}|$, the spatiotemporal amplitude summation decreases asymmetrically, as typically expected[6]. On the other hand, we also study the statistics of the synaptic output when two spikes with different power levels are simultaneously applied to the device (see Supplementary Fig. S6). Briefly, we compare the measured and theoretical arithmetic sum of the two separated IPSCs. It is found that close-to-linear spatial summation applies in the low power regime (where IPSC does not saturate), although this transits to a sublinear-dependence by increasing either spikes' power into the saturation regime. It is worth mentioning that the observed sub-linear summation rule is analogous to the action of a soma cell[34].

Finally, the rich physics associated with hot-carriers in graphene and charge transfer dynamics at the graphene-SWNTs interface make our device a unique platform for achieving complex functions and computing capabilities. To demonstrate optical processing on light stimuli, we employ one of the light channels (405 nm spike B) to conceptually emulate a neuromorphic feedback signal and demonstrate optical AND/OR logic operation on the stimuli (405 nm spike A) through charge-trap-mediated optical coupling. It is worth pointing out that we choose to demonstrate logic operations due to its straightforwardness. In principle, the function generation can be readily extended to neuromorphic tasks. Specifically, AND logic operation is performed based on the super-linear temporal summation rule. When the gate bias is set close to $V_{cross}$, a very low output current is obtained under individual pre-synapse stimulation. However, if both pre-synapses trigger simultaneously, super-linear summation rule would guarantee a significant output, as shown in Fig. 4c. On the other hand, sub-linear power summation rule results in the OR operation. Setting the gate bias at $V_G=20$ V, the device is operating in the IPSC saturation regime, where any pre-synaptic pulse would trigger a high output current. Sub-linear power summation rule ensures that the output current does not change much even both pulses are applied. OR logic operation is therefore implemented, as plotted in Fig. 4d. It should be noted that effective coupling between different light fields in our device opens up a plethora of possibilities to further optimize the device performance. By engineering our device structure to allow coupling with evanescent light fields in a waveguide, it is even feasible to realize fully integrated photonic neuromorphic systems[13, 35].

**Methods**

The graphene samples were grown on copper foil by CVD method, and Raman spectroscopy combined with optical microscope characterizations point to a defect-free single-layer sample (Supplementary Section S6). The single-wall carbon nanotubes (SWNTs) were purchased from a commercial supplier (Carbon solutions Inc.). To fabricate a proof-of-concept artificial synapse, SWNT suspensions are produced by ultrasonicating 2 mg nanotube in 20 mL NMP, and then the resulting suspensions ultra-centrifuged with 10,000 g for 1h was coated onto a Si/SiO$_2$ (285 nm) wafer. CVD graphene is transferred on top of the SWNT layer using the poly(methyl methacrylate) supported procedures. Subsequently, different metal composition (Ti/Au and Pd/Au) as the source and drain electrodes are patterned by standard photolithography. Graphene channel fabrication is patterned by another photolithography and oxygen plasma etching. For simulating synaptic functions, we employ 405 and 532 nm laser diodes as pre-synapse spikes, respectively. The beam is guided through an optical fiber with a FC/PC ferrule and is subsequently incident onto the channel of the devices without focusing. The beam at the device was measured to be Gaussian-shaped with a diameter of about 300 μm (at 405 nm illumination). The area of the channel is about 90 μm×30 μm.

Electrical performance of the transistors and synaptic simulation measurements are carried out by a semiconductor parameter characterization system (Keithley 4200 SCS) in a closed probe station under vacuum ($10^{-6}$ Torr) at room temperature. Pre-synaptic light spikes are applied on the graphene channel, and post-synaptic current is measured by applying a source-drain voltage of 0.5 V. To verify the universality of the operation of our optical synapses, excitatory/inhibitory post-synaptic currents and PPF stimulated by spikes from five synaptic devices are measured.

Raman measurements were performed in a Horiba Jobin Yvon LabRAM HR 800 system using a 514 nm excitation laser operating at 1 mW, ×100 objective lens with about 1 μm diameter spot size, and 1800 lines/mm grating with about 0.45 cm$^{-1}$ spectral resolution.

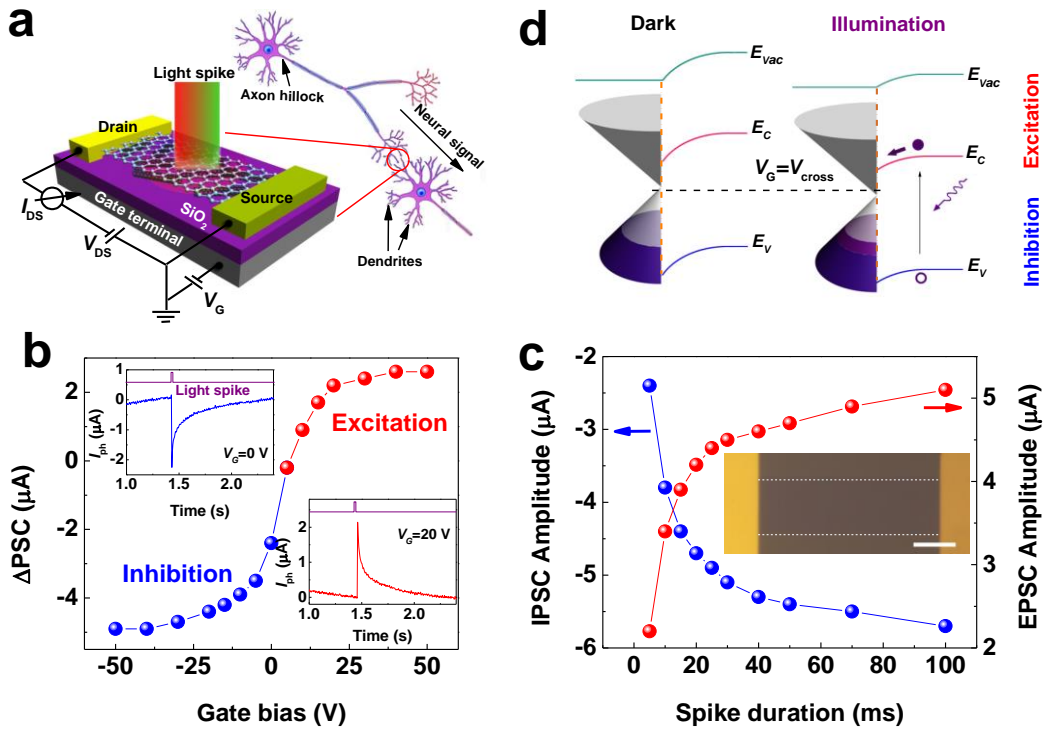

**Figure 1. Light-activated synapse with *in-situ* adjustment function. a,** Schematic illustrations of the synapse and the neural signal transmission of a biological synaptic integration in a neuron. In a biological synapse, action potentials caused by firing permits neurotransmitter release, then the signals are transmitted as a synaptic potential from the pre-synapse to the post-synapse. In the case of our artificial synapse, application of the light pulse generates several photocarriers, which contribute to the current change in the channel. **b,** The change of amplitudes of PSC triggered by a pre-synaptic light spike (50 μW, 5 ms) at gate biases between -50 V and +50 V. Top inset: A typical IPSC change triggered by the light spike under $V_G=0$ V. Bottom inset: a typical EPSC change in response to the light spike under $V_G=20$ V. **c,** Spike duration-dependent change of IPSC and EPSC for an inhibitory synapse (blue balls line, $V_G=0$ V) and an excitatory synapse (red balls line, $V_G=20$ V), the power of the light spike is 50 μW. The inset is the optical micrograph of the fabricated device (Scale bar, 20 μm). **d**, Energy-band diagram of the graphene/SWNTs interface. The built-in field formed at the interface is tuned by electrons transferred from the SWNTs layers in illumination. Purple dot, circle and arrow indicate electron, hole and photon, respectively.

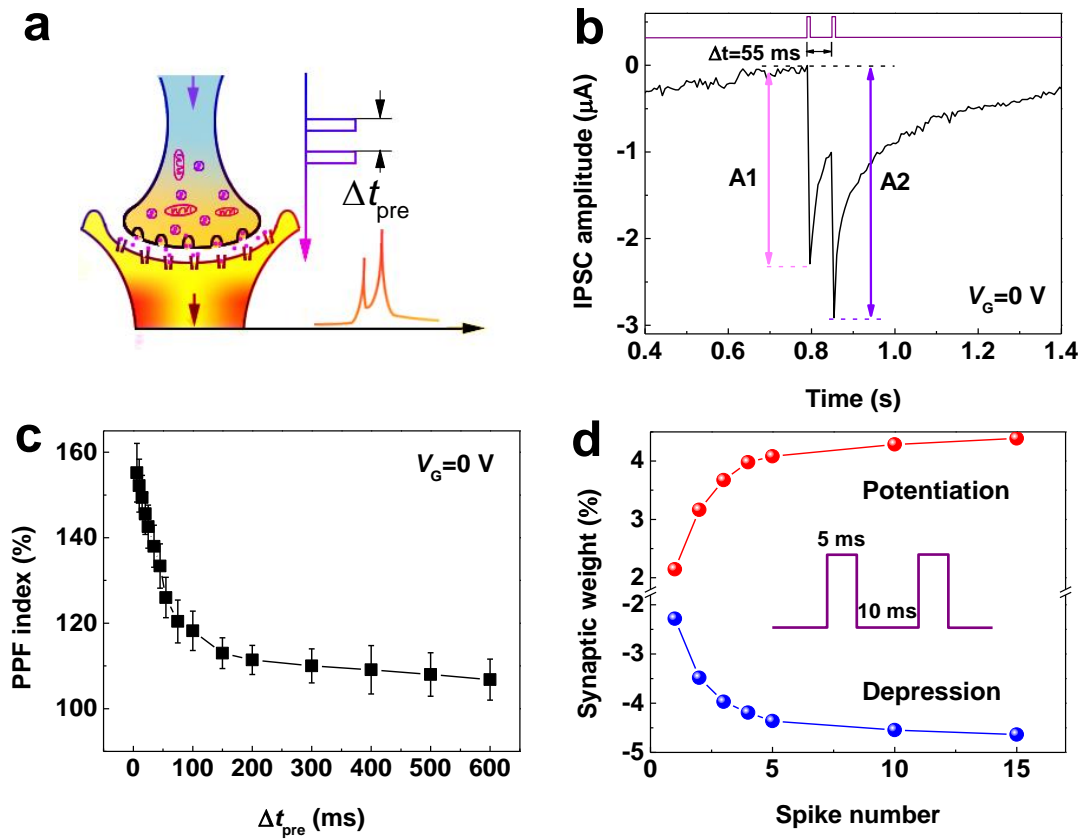

**Figure 2. PPF behaviors by temporally correlated light spikes. a**, Schematics of the post-synaptic current triggered by a pair of temporally correlated pre-synaptic spikes. **b**, The change of amplitude of IPSC by a pair of pre-synaptic light spikes with an inter-spike interval time of 55 ms at $V_G$=0 V. A1 and A2 represent the amplitudes of the first and second IPSCs, respectively. **c**, PPF index (A2/A1) plotted as a function of light spike interval time $\Delta t_{pre}$, which is extracted from 10 independent tests. The dots and error bars represented the average values and standard deviations, respectively. **d**, The depression and potentiation modulation for the inhibitory (blue balls line, $V_G$=0 V) and excitatory (red balls line, $V_G$=20 V) synapses operated by a series of consecutive light spikes (50 μW, 5 ms pulses with 10 ms intervals).

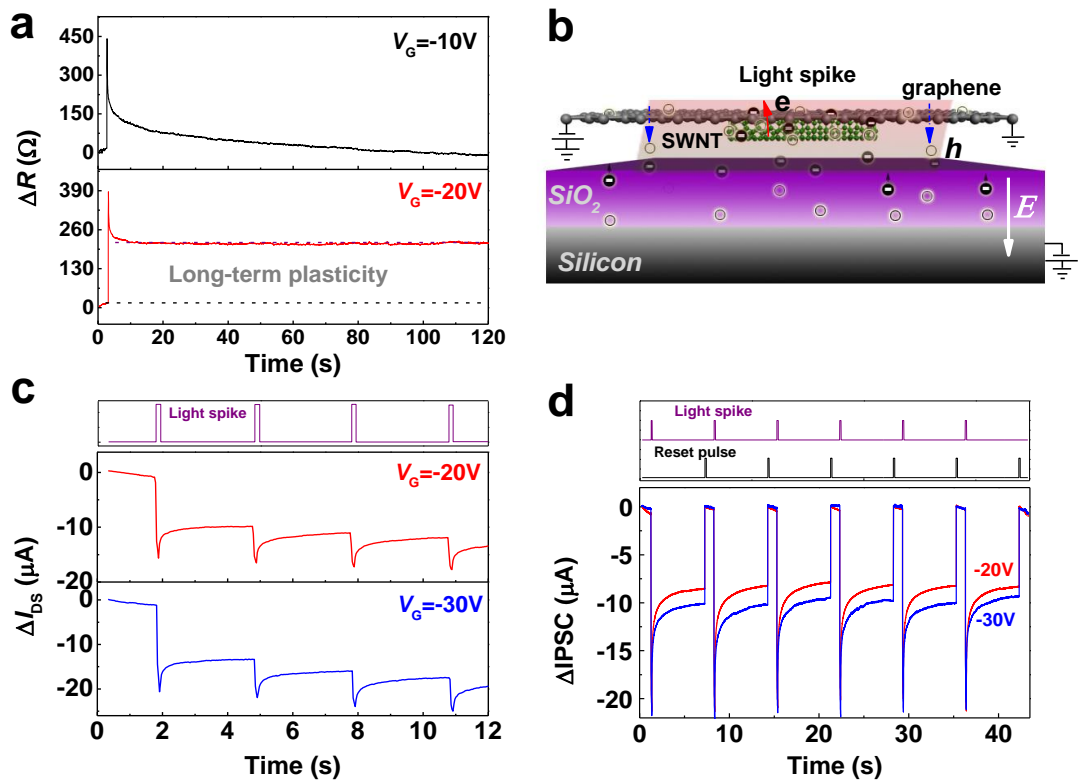

**Figure 3. Gate-dependent LTP formation in the artificial synapse. a**, The amplitude of IPSC triggered by a long-duration pre-synaptic light spike (50 μW, 100 ms) at $V_G$=-10 V and -20 V, respectively. **b**, Schematic illustration of LTP formation mechanism, photo-generated holes from the graphene FET channel or SWNTs are captured at the surface of silicon dioxide due to the existence of the defects. The white arrow indicates the direction of the electric field for negative gate voltages. **c**, The amplitude of IPSCs triggered by a serial of consecutive spikes (100 ms pulses spaced 3 s apart with 50 μW average power) at different gate bias, showing enhancement of LTP effect. **d**, Gate voltage-dependent amplitudes of IPSC and reset switching operations by the gate bias pulses.

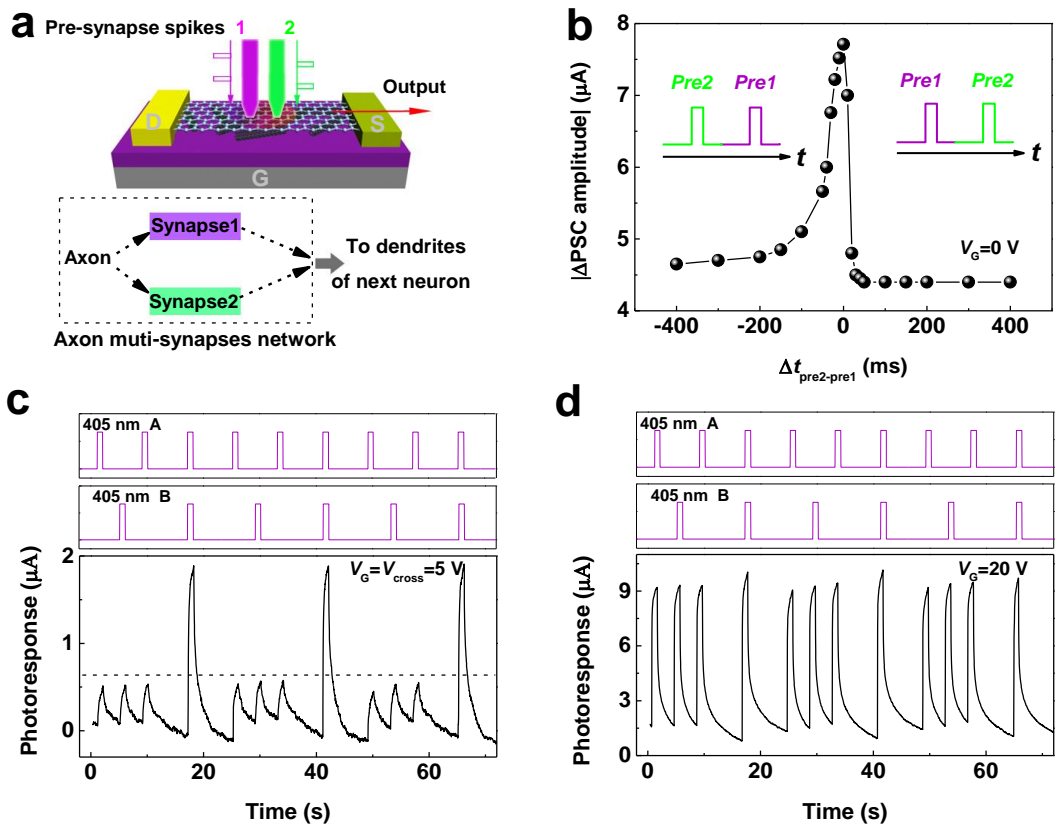

**Figure 4. Artificial axon multi-synapses network and optical logic operations. a**, Schematic illustrations of the multi-neural signals transmission in a biological system and a compact artificial axon-multi-synapses network employing multiple light stimuli. **b**, The change of PSC amplitude at the trailing edge of pre-synapse1 spike as a function of $\Delta t_{pre2\text{-}pre1}$ between the two presynaptic spikes (duration 20 ms with different power for 405 nm and 532 nm spikes) at $V_G$=0 V. **c**, The time response of an AND logic operation by utilizing two 405 nm light spikes (10 μW, 1 s) as logic input signals at $V_{cross}$ (~5 V in this irradiation operation). **d**, Demonstration of the OR logic operation at $V_G$=20 V when the pre-synapse spikes are chosen as 50 μW, 1 s.

*Correspondence emails:

Fengqiu Wang - fwang@nju.edu.cn;

Xiaomu Wang - xiaomu.wang@nju.edu.cn;

Rong Zhang - rzhang@nju.edu.cn;